# Analytic Methods to Calculate Fault Trees with Loops - Restrictions of Application and Solution Uniqueness

SERGEY POROTSKY

**Abstract**. One of the important tasks of the Reliability Estimation is Analysis of the Fault Tree. A problem of Fault Trees analysis is considered one of the most complex ones, since structure of such trees is characterized by a considerable number of interconnections. Usually analytical methods are used and most applicable method is Minimal Cut Sets building and calculation. Classical Fault Tree Analysis methods are applicable only for Fault Trees without loops. Loops can appear in Fault Tree, when a TOP or some intermediate gates appear as input to another gate at a lower level of the model. An occurrence of a Loop has been a problematic issue in a Fault Tree calculation.

The article relates to the uniqueness of the solution for the Fault Trees with arbitrary Loops. There are assumed, that failures of the Basic Events are non-repairable and Fault Tree gates may be expressed by two main logic gates – **AND**-gates and **OR**-gates.

## 1. INTRODUCTION

One of the important tasks of the Reliability Estimation is Analysis of the Fault Tree [1]. Classical Fault Tree Analysis methods (Minimal Cut Sets calculations) are applicable only for Fault Trees **without loops**. A variety of methods have been developed to calculate Fault Tree with loops (see, e.g., articles [1 – 9]). Fault Tree Handbook with Aerospace Applications [1] proposes general advise – "…The loops are cut (eliminated) in the fault tree…". But it is non-correct to simply "delete loop", analyst should carefully investigate concrete features of the analyzed Fault Tree and to decide, how to build equivalent Fault tree without loop.

The conventional method, presented on [2], proposes to solve the logical loop problem by breaking the logical loops at the points where the dependencies among the support systems are relatively weak and developing new fault trees without the logical loops. But this method gets us exact solution only for simple FaultTrees with loops. Yang [3] built contra-example for this approach, which shows its mistake. Consider Fault Tree of 4 TOPs and triple linear interrelated loops:

$$A = Aa \text{ OR } (Ab \text{ AND } B) \text{ OR } (Ac \text{ AND } C) \text{ OR } (Ad \text{ AND } D)$$
$$B = Bb \text{ OR } (Ba \text{ AND } A) \text{ OR } (Bc \text{ AND } C) \text{ OR } (Bd \text{ AND } D)$$
$$C = Cc \text{ OR } (Ca \text{ AND } A) \text{ OR } (Cb \text{ AND } B) \text{ OR } (Cd \text{ AND } D)$$
$$D = Dd \text{ OR } (Da \text{ AND } A) \text{ OR } (Db \text{ AND } B) \text{ OR } (Dc \text{ AND } C)$$

Where:

- A, B, C and D are Sub-Fault Trees TOPs with Loops, non-calculated directly by means of classical non-cycled fault tree analysis
- Aa, Ab, Ac, Ad, Bb, Ba, Bc, Bd, Cc, Ca, Cb, Cd, Dd, Da, Db and Dc are Basic Events.

On the [3] it is shown, that for Ad = Dc = Cb = Db = **TRUE** and other Basic Event values, equalled for **FALSE**, the value of TOP, obtained by conventional method using, is equalled to **FALSE**, but it isn't satisfy for above equations. Opposite, value
A = **TRUE** is correct.

Yang [3] presented an exact analytical method to break the logical loops by means of using of the Boolean Algebra laws to transformate Fault Trees with loops to the Fault Tree without loops. It is proposed to break the logical loops in the merged fault tree by disconnecting one of the connected gates that cause the logical loops. Some modifications of this approach are considered in different articles, denoted for analysis of the Fault Tree with loops [5 – 7].

Proposed of these articles methods have following drawbacks:
- They don't formulate restrictions for its field of application. Arbitrary Fault Trees may consist of gates of different types (**AND**, **OR**, **NOT**, etc.) and Basic Events of different types (non-repairable, repairable, periodically tested, etc.). Really these articles consider only Fault Trees with gates **AND** and **OR**, but assumptions according Basic Event types are absent. Otherwise, for repairable Basic Events proposed methods don't get correct results.
- It isn't proved, that proposed analytic methods get us full solution, i.e. don't exist other Boolean solutions, also satisfy for analysed Fault Tree with loops. These methods use direct Boolean transformation, so they are not applicable for situation, when simultaneously two possible TOP values (both **FALSE** and **TRUE**) are satisfy for the Boolean equations and so may be solutions.

The main aim of theour article is to remove these drawbacks of the early proposed analytic methods. Remaining part of the paper is organized as follows:
- In Chapter 2 we introduce some contra-example for analytic methods, proposed for Fault Trees with Loops
- In Chapter 3 we prove Uniqueness of the solution for Arbitrary Fault Trees with Loops and Non-Repairable Basic Events
- In Chapter 4 we discuss obtained results.

## 2. FIELD of APPLICATION for ANALYTIC SOLUTIONS

### 2.1 Types of Basic Events

Consider following Fault Tree:

$$A = Aa \textbf{ OR } (Ab \textbf{ AND } B),$$
$$B = Bb \textbf{ OR } (Ba \textbf{ AND } A)$$

Where A and B are TOPs, Aa is repairable Basic Event, Ab, Bb and Ba are are repairable or non-repairable Basic Events.

According analytic methods, proposed on [3 - 9] the solution will be following:

- A = Aa **OR** (Ab **AND** Bb)

So, for combination {Aa = **FALSE,** Ab = **TRUE,** Ba = **TRUE**, Bb = **FALSE** } we get solution A = **FALSE.**

Consider following trajectories of the Basic Event state transformations and TOP A state transformation, due to equation A = Aa **OR** ( Ab **AND** ( Bb **OR** (Ba **AND** A) ) ) :

$t_0$: Aa($t_0$) =**FALSE**, Ab($t_0$) =**FALSE**, Ba($t_0$) =**FALSE,** Bb($t_0$) =**FALSE,** A($t_0$) =**FALSE**
$t_1$: Aa($t_1$) =**TRUE**, Ab($t_1$) =**FALSE**, Ba($t_1$) =**FALSE,** Bb($t_1$) =**FALSE,** so A($t_1$)=**TRUE**
$t_2$: Aa($t_2$) =**TRUE**, Ab($t_2$) =**TRUE**, Ba($t_2$) = **FALSE,** Bb($t_2$) =**FALSE,** so A($t_2$)=**TRUE**
$t_3$: Aa($t_3$) =**TRUE**, Ab($t_3$) =**TRUE**, Ba($t_3$) = **TRUE**, Bb($t_3$) = **FALSE**, so A($t_3$) = **TRUE**
$t$:  Aa($t$) = **FALSE**, Ab($t$) = **TRUE**, Ba($t$) = **TRUE,**  Bb($t$) = **FALSE,**   so A($t$) = **TRUE**

where $t_0$ is Initial Time, $t_1$, $t_2$ and $t_3$ are some Intermediate Times, $t$ is Final Time (i.e. the time of the Fault Tree Analysis)

Is it mean, that solution, obtained by analytic methods proposed on [3 - 9], is wrong? No, this solution also may be, but for another trajectory:

$t_0$: Aa($t_0$) =**FALSE**, Ab($t_0$) =**FALSE**, Ba($t_0$) =**FALSE,** Bb($t_0$) =**FALSE,** A($t_0$) = **FALSE**
$t_1$: Aa($t_1$)=**FALSE**, Ab($t_1$) =**TRUE**, Ba($t_1$) =**FALSE,** Bb($t_1$) =**FALSE,** so A($t_1$)=**FALSE**
$t_2$: Aa($t_2$)=**FALSE**, Ab($t_2$) =**TRUE**, Ba($t_2$) =**TRUE,** Bb($t_2$) =**FALSE,** so A($t_2$) =**FALSE**
$t$:  Aa($t$) = **FALSE**, Ab($t$) = **TRUE**, Ba($t$) = **TRUE,**  Bb($t$) = **FALSE,** so A($t$) = **FALSE**

Due to possible non-monotonically character of the Basic Event Aa trajectoty (**FALSE→TRUE→FALSE),** we get dual solutions of the TOP for the same initial and final values of the Basic Events. So, for Fault Trees with Loops and Repairable Basic Events the solution depends not only of values of Basic Events, but also of Basic Event state transformation, i.e. of pre-history of Basic Events States. Same conclusion is correct for some other types of Basic Events – periodically tested, constant availability, etc. But for Fault Trees with Loops, which contain only Non-Repairable Basic Events, this situation (dual solution for some combination of the Basic Event values) is impossible, we will prove this in next chapter.

***Comment.*** For first point of view, this fact (existence of the dual solutions) is some strange. Classic Fault Tree Analysis methods (Minimal Cut Sets) don't take into account pre-history of the Basic Event states and allow us to Calculate TOP state based only of the final states of the Basic Events, both for repairable and non-repairable Basic Events. But it is correct only for classic, regular Fault Tree. For some other types of the Fault Trees it is wrong. For example, for Dynamic Fault Trees, which contain such gates as **PAND (Priority AND)**, **SEQ, SPARE, FDEP,** etc., it is necessary to take into account pre-history of the single gate state transformations – final states of the Basic Events don't allow us to calculate TOP state.

## 2.2 Types of Gates

If Fault Tree contain some NOT type gates (**NOT, NOR, NAND, XOR,** etc.**)** , we also could not provide monotonicall character of the some gate trajectoty and so dual solutions may exist.

So, second (aFault Treeer Basic Event types) restriction should be following: all gates on the analysed Fault Tree may be expressed by two main logic gates – **AND**-gates and **OR**-gates.

Certainly, Fault Tree can contain some complex gates, e.g "K-out-of-N" gate, but should be possible expressed these complex gates only from **AND**-gates and **OR**-gates. For example, "2-out-of-3" gate may be expressed as:

{Gate1 **AND** Gate2 **AND** **NOT**(Gate3) } **OR** {Gate1 **AND** Gate3 **AND** **NOT**(Gate2) } **OR** {Gate2 **AND** Gate3 **AND** **NOT**(Gate1) } **OR** {Gate1 **AND** Gate2 **AND** Gate3}**,** but this expession contains not only **AND**-gates and **OR**-gates, but also **NOT** gate and so directly not applicable for analytic methods, proposed for Fault Trees with Loops. But it is also possible to expressed "2-out-of-3" gate only from **AND**-gates and **OR**-gates, without using of gate **NOT**:

{ Gate1 **AND** Gate2 } **OR** { Gate1 **AND** Gate3 } **OR** { Gate2 **AND** Gate3 }, and this Fault Tree applicable for analytic methods, proposed for Fault Trees with Loops.

## 3. Uniqueness of the solution

**Main Theorem.**

*For arbitrary Fault Tree with multiple non-linear interrelated Loops, which contains only non-repairable Basic Events and all gates may be expressed only by means of AND-gates and OR-gates, for possible dual solution of some TOP all trajerctories will get same output values of this TOP.*

### 3.1 FAULT TREES with ORDINARY LOOPS

First consider Fault Tree, for which the loops are only ordinary. Ordinary loop means, that each TOP may depend step-by-step (by circle) only of one other TOP. Full Fault Tree may consist on several sub-trees and the TOP of each sub-tree depends not only of Basic Events, but also from one other TOP:

TOP[1] = F(a[1],…, a[n], TOP[2])

TOP[2] = U(a[1],…, a[n], TOP[3])

............................

TOP[k-1] = W(a[1],…, a[n], TOP[k])

TOP[k] = R(a[1],…, a[n], TOP[1])

Where:

- F, U,…,W, R – some boolean expressions.
- a[1],…, a[n] – Basic Events

Illustration of dependencies between fault trees:

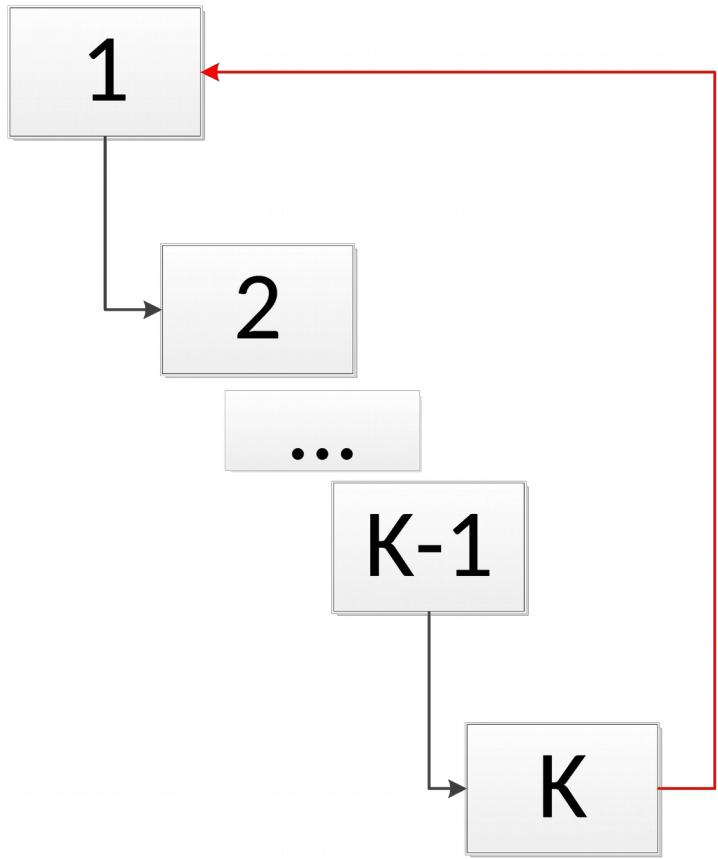

We can sequentially substitute expressions for TOP and to get final equation

(1)  TOP = H(a[1],…, a[n], TOP),

where H is some boolean expression and for TOP[1] the index is omitted (i.e.

TOP = TOP[1] ).

Using standard rules of the Fault Tree Minimal Cut building, following representation always may be got:
TOP = {a[$i_1$] **AND** a[$i_2$] **AND** …} **OR** {a[$j_1$] **AND** a[$j_2$] **AND** …} **OR…OR** {a[$m_1$] **AND** a[$m_2$] **AND** …} **OR** {TOP **AND** a[$q_1$] **AND** a[$q_2$] **AND** …} **OR…OR** {TOP **AND** a[$f_1$] **AND** a[$f_2$] **AND** …}.

From this representation, short form may be get (***MAIN EQUATION***):

(2)         TOP = $Q_{10}$(a[1],…, a[n]) **OR** {$Q_{11}$(a[1],…, a[n]) **AND** TOP}

Where $Q_{10}$ (a[1],…, a[n]) and $Q_{11}$(a[1],…, a[n]) – some boolean expressions, which don't depend of TOP and depend only of Basic Events a[1],…, a[n] and Fault Tree structure.
All state alternatives are shown on the Table 1, and column "Available" corresponds for the possible correct solution ( " + " corresponds for available value, i.e. if "input TOP value" = "Top value from eq. (2), " - " corresponds for non-available value, i.e. if "input TOP value" =/= "Top value from eq. (2) ). Possible Dual Solutions are signed as ***Italic&Bold***

| Values of the input variables at the final time t | | | Output Value of TOP at the final time t | | | |
|---|---|---|---|---|---|---|
| $Q_{10}$ | $Q_{11}$ | TOP value | From eq. (2) | TOP value is Available ? | By means of simulation | Correct value |
| FALSE | FALSE | FALSE | FALSE | + | | FALSE |
| | | TRUE | FALSE | - | | |
| FALSE | TRUE | ***FALSE*** | FALSE | + | FALSE | FALSE |
| | | ***TRUE*** | TRUE | + | | |
| TRUE | FALSE | FALSE | TRUE | - | | TRUE |
| | | TRUE | TRUE | + | | |
| TRUE | TRUE | FALSE | TRUE | - | | TRUE |
| | | TRUE | TRUE | + | | |

Table 1. States alternatives for arbitrary Fault Tree with ordinary loop

It is seen, that from 4 alternatives for $Q_{10}$ and $Q_{11}$ values, only one combination {$Q_{10}$ = **FALSE,** $Q_{11}$ = **TRUE**} gets possible dual solution: both TOP = **TRUE** and TOP = **FALSE** may be solutions of the main equation (2). For this situation it is necessary to perform event-driven (step-by-step) simulation.

Assume, that all Basic Events a[i] have value **FALSE** as the initial state. If some a[i]($t_0$) = **TRUE,** due to non-repairable type of the Basic Events it will be also same value on this state at the final time t (i.e. a[i](t) = **TRUE**), so it is necessary to perform following:
- To remove Basic Event a[i] from the gates **AND,** for which a[i] is input.
- To install **TRUE** instead of output of the gate **OR,** for which Basic Event a[i] is input, and etc, follow to Down-Top approach.

As at the start (i.e. at the time $t_0$) all Basic Events are at the state **FALSE**, the $Q_{10}(t_0)$ and $Q_{11}(t_0)$ are also have values **FALSE,** because they are some combinations of the **AND-**gates and **OR-**gates. So, TOP($t_0$) is also has value **FALSE**, to satisfy eq. (2). Consider some intermediate time moment $t_1$ s.t. $t_0 < t_1 < t$, at what some Basic Event a(i) has changed his state from **FALSE** to **TRUE** (if ALL Basic Events did not change its states up time t, TOP also did not change its state and so TOP(t) = TOP($t_0$) ).
    Is it possible, that value of $Q_{10}(t_1)$ = **TRUE** ? No, it is impossible, because all Basic Events are non-repairable and so any Boolean Function, composed of gates **AND** and **OR,** is non-decreased and at the time moment t the Boolean Function $Q_{10}(t)$ should have value **FALSE**.

Is it possible, that value of $Q_{11}(t_1)$ = **TRUE** ? Yes, it is possible, because at the time moment t the Boolean Function $Q_{11}(t)$ should have value **TRUE**. Also possible, that $Q_{11}(t_1)$ = **FALSE.**

But in any case, independently of $Q_{11}(t_1)$ value (**TRUE** or **FALSE**)**,** the new value of the the TOP at the time $t_2 = (t_1 + \Delta t$ ) will be **FALSE,** because TOP($t_2$) = $Q_{10}(t_1)$ **OR** {$Q_{11}(t_1)$ **AND** TOP($t_1$)} = **FALSE OR** {$Q_{11}(t)$ **AND FALSE**)} = **FALSE**.

So, after each possible state changing of some Basic Event from **FALSE** to **TRUE** the state of the TOP isn't changed and will have same value **FALSE**.

***Conclusion.*** Analysis of the Table 1 allow us to get equation TOP(t) = $Q_{10}(t)$ and so to produce following rule (named "**FALSE insertion in the initial Fault Tree with loop instead of TOP input**") - to calculate Fault Tree with ordinary loops it is enough to delete loops and to insert to the right part of the initial equation (2) value **FALSE** as Input instead of the TOP input**.**

### 3.2. ARBITRARY FAULT TREES

Consider now arbitrary Fault Tree, for which the loops are not only ordinary and moreover, are not only linear. For example, Fault Tree with 3 non-linear interrelated loops is following:

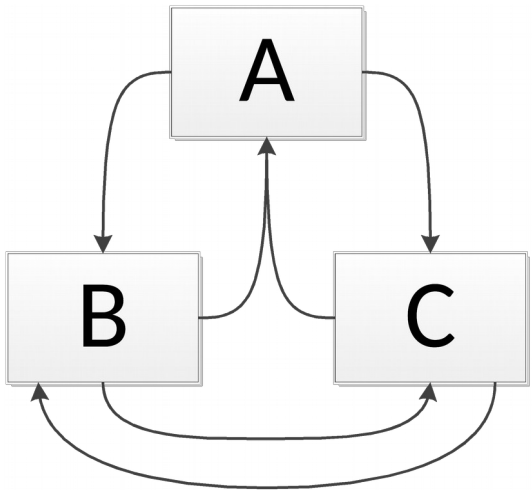

(3)
$$A = Aa \textbf{ OR } (Ab \textbf{ AND } B) \textbf{ OR } (Ac \textbf{ AND } C) \textbf{ OR } (Abc \textbf{ AND } B \textbf{ AND } C)$$
$$B = Bb \textbf{ OR } (Ba \textbf{ AND } A) \textbf{ OR } (Bc \textbf{ AND } C) \textbf{ OR } (Bac \textbf{ AND } A \textbf{ AND } C)$$
$$C = Cc \textbf{ OR } (Ca \textbf{ AND } A) \textbf{ OR } (Cb \textbf{ AND } B) \textbf{ OR } (Cab \textbf{ AND } A \textbf{ AND } B)$$

where A, B and C are TOPs of the Sub-Fault Trees with loops.
By means of substitution of C to the first and second equation, and Boolean algebra transformations, we get following main equations (4):

(4)    $A = Q_{10}$ **OR** ($Q_{11}$ **AND** A) **OR** ($Q_{12}$ **AND** B)  **OR** ($Q_{13}$ **AND** A **AND** B)

   $B = Q_{20}$ **OR** ($Q_{21}$ **AND** A) **OR** ($Q_{22}$ **AND** B)  **OR** ($Q_{23}$ **AND** A **AND** B)

where $Q_{10}$, $Q_{11}$, etc. – some boolean expressions, dependent only from Basic Events a[1],…, a[n], and don't dependent from TOPs A and B.

For arbitrary Fault Trees with multiple loops, where each TOP may depend Non-Linearly of several TOPs, the system of the ***main equations*** for the TOP[i] will be following

(i = 1…k):

(5) $\quad$ TOP[i] = $Q_{i,0}$ **OR** {($Q_{i,1}$ **AND** TOP[1])} **OR** … **OR** {($Q_{i,k}$ **AND** TOP[k])} **OR** {($Q_{i,1,2}$ **AND** TOP[1] **AND** TOP[2])} **OR** …**OR** {($Q_{i,1,k}$ **AND** TOP[1] **AND** TOP[k])} **OR** …**OR** {($Q_{i,1,2,…,k}$ **AND** TOP[1] **AND** TOP[2] **AND**…**AND** TOP[k])}

Where $Q_{i,j,…}$(a[1],…, a[n]) – some Boolean expressions (MinCut Sets), depending on Basic Events a[1],…, a[n], and don't dependent on TOP[1],,,TOP[k].

Fragment of the full table for arbitrary Fault Tree with 3 TOPs and triple non-linear interrelated loops, corresponded for the system of equations (4), is shown on the Table 2. Column "Available" corresponds for the possible correct solutions, possible dual solutions are signed as ***Italic&Bold***.

As for Fault Trees with ordinary loops it is possible to assume, that at the moment $t_0$ the values of all Basic Events a[1],…, a[n] are equalled to **FALSE.** It is also clear, that values of all Boolean Expressions $Q_{i,j,k}(t_0)$ are also equalled to **FALSE** and so, to satisfy eq. (5), values of all $T(t_0)$ should be also equalled to **FALSE**.

| | Values of the Input Variables at the final time t | | | | | | | | | | Output values of A and B at the final time t | | | |
|---|---|---|---|---|---|---|---|---|---|---|---|---|---|---|
| | | | | | | | | | | | From eq. (4) | | Values are available? |
| | $Q_{10}$ | $Q_{11}$ | $Q_{12}$ | $Q_{13}$ | $Q_{20}$ | $Q_{21}$ | $Q_{22}$ | $Q_{23}$ | A | B | A | B | |
| 1 | False | False | False | False | False | False | False | False | False | False | False | False | + |
| | | | | | | | | | False | True | False | False | - |
| | | | | | | | | | True | False | False | False | - |
| | | | | | | | | | True | True | False | False | - |
| 2 | False | False | False | False | True | False | False | False | False | False | False | True | - |
| | | | | | | | | | *False* | True | False | True | + |
| | | | | | | | | | True | False | False | True | - |
| | | | | | | | | | *True* | True | True | True | + |
| 3 | False | False | True | False | False | True | False | False | *False* | *False* | False | False | + |
| | | | | | | | | | False | True | False | False | - |
| | | | | | | | | | True | False | False | False | - |
| | | | | | | | | | *True* | *True* | True | True | + |

Table 2. Fragment of the full table of the states alternatives for arbitrary Fault Tree with 3 TOPs and triple non-linear interrelated loops.

**Statement 1.** For any intermediate time moment any TOP[i] could not change his state from **TRUE** to **FALSE.**

To prove this statement, consider some time moment $t_1$, at what TOP[i] = **TRUE** and some Basic Event a[j] has changed his state (if all Basic Events did not change its states, each TOP also did not change its state). Due to non-repairable type of the Basic Events, the only transformation of Basic Event state from **FALSE** to **TRUE** is possible, i.e. a[j]($t_1$) = **TRUE**. Consider eq. (5) – it consist only of Boolean operations **AND** and **OR,** and so for time $t_2 = t_1 + \Delta t$ we get, that TOP[i]($t_2$) can't decrease its value for comparison with TOP[i]($t_1$) value, so TOP[i]($t_2$) = **TRUE**

Consider some fixed combination of the possible values of the Basic Events at the time t {a[1](t),…, a[n](t)} and corresponding combination of values of Boolean expressions $Q_{i,j,..}( a[1](t),…, a[n](t) )$. Consider some possible dual solution for some TOP[i](t) according eq. (5). As illustration we can consider values **FALSE** and **TRUE** for TOP = A from Fault Tree (4) – see lines 2 and 3 from Table 2. There may be two alternatives:

a) For this combination of values of the Boolean expressions $Q_{i,j,..}(a[1](t),…, a[n](t)$ only one TOP has possible dual solution ( e.g., line 2 at the Table 2 – only A has possible dual solution)

b) For this combination of values of the Boolean expressions $Q_{i,j,..}(a[1](t),…, a[n](t) )$ both several TOPs simultaneously have possible dual solutions (e.g., line 3 at the Table 2 – both A and B have possible dual solutions)

a) Let only TOP[i] has possible dual solution for this combination of the Boolean expressions $Q_{i,j,..}(a[1](t),…, a[n](t))$. From eq. (5) we get, that $Q_{i,0}(t)$ = **FALSE**, because in opposite case TOP[i](t) = **TRUE** for any possible values of the $Q_{i,j,..}(t)$ and TOP[j](t) and so solution for TOP[i](t) will not be possible dual.

***Comment***. It is right only for TOPs with possible dual solutions. For example, for B we see, that $Q_{20}(t)$ = **TRUE** (line 2 at the Table 2)

After separation of all expressions on the right part of the equation (5) for TOP[i] between expressions, which contain TOP[i] , and expressions, which don't contain TOP[i], we can re-write eq. (5) in shortest form as :

(6)   TOP[i] = $Q_{i,0}$ **OR** $G_i$ **OR** ($W_i$ **AND** TOP[i]),

Where:

- $G_i$ – some Boolean expression, that don't dependent on TOP[i].

- $W_i$ – some Boolean expression, that dependent or don't dependent on TOP[i].

For considered combination of the Boolean expressions $Q_{i,j,..}( a[1](t),…, a[n](t) )$ the only TOP[i](t) has possible dual solution, so all other TOPs have some fixed (single) solutions. If for these solutions and for considered combination of the Boolean expressions $Q_{i,j,..}(a[1](t),…, a[n](t))$ we have, that $G_i(t)$ = **TRUE,** we get that  TOP[i](t) = **TRUE** for any values of the $W_i(t)$ and TOP[i](t), and so solution for TOP[i](t) will not be possible dual. So, $G_i(t)$ = **FALSE.**

Consider some intermediate time moment $t_1$ s.t. $t_0 <= t_1 <= t$, at what some BE has changed his state from **FALSE** to **TRUE** (if all BEs did not change its states up time t, all Boolean expressions $Q_{i,j,..}(a[1],…, a[n])$ also did not change its state, and so only TOP[i](t) = **FALSE** will be solution and we haven't dual solution – see line 1 at the Table 2).

Both all BEs and all TOPs could not change its states from **TRUE** to **FALSE,** so boolean expressions $Q_{i,0}$ and $G_i$ also could not change its states from **TRUE** to **FALSE** (because they composed only from operations **AND** and **OR**).

So, at the time $t_1$ for $Q_{i,0}$ and $G_i$ there are proved, that $Q_{i,0}(t_1)$ = **FALSE** and $G_i(t_1)$ = **FALSE**, because both $Q_{i,0}(t_0) = Q_{i,0}(t)$ = **FALSE** and $G_i(t_0) = G_i(t)$ = **FALSE**.

At the time $t_1$, when some BE has changed his state from **FALSE** to **TRUE**, the value of the TOP[i] is as early, i.e. TOP[i]($t_1$) = **FALSE**. We don't know value of the $W_i(t_1)$ - it may be both **TRUE** and **FALSE**. But in any case at the time $t_2 = t_1 + \Delta t$ (after some of the Basic Events has changed its state from **FALSE** to **TRUE**) the value of the TOP[i] at this time $t_2$ will be :

TOP[i]($t_2$) = $Q_{i,0}(t_1)$ **OR** $G_i(t_1)$ **OR** {$W_i(t_1)$ **AND** TOP[i]($t_1$) } = **FALSE OR FALSE OR** {$W_i(t_1)$ **AND FALSE**) } = **FALSE**.

So, after each transformation of the state of any of Basic Events the value of the TOP[i] with dual solution isn't changed and so TOP[i](t) = **FALSE.**

b) For some combination of values of the Boolean expressions $Q_{i,j,..}(a[1](t),…, a[n])(t)$ both several TOPs simultaneously have dual solutions ( e.g., line 3 at the Table 2 – both A and B have dual solutions).

Let TOP[i] and TOP[j] have dual solutions for this combination of the Boolean expressions $Q_{i,j,..}(a[1](t),…, a[n](t))$. From eq. (5) we get, that $Q_{i,0}(t)$ = **FALSE** and $Q_{j,0}(t)$ = **FALSE**, because in opposite case TOP[i](t) = **TRUE** or TOP[j](t) = **TRUE** for any possible values of the $Q_{i,j,..}(t)$ and TOP[r](t) and solutions for TOP[i](t) and TOP[j](t) will not be dual.

After separation of all expressions on the right part of the equation (5) for TOP[i] between expressions, which contain TOP[i] or TOP[j] , and expressions, which don't contain TOP[i] or TOP[j], we can re-write eq. (5) in shortest form as :

(7)   TOP[i] = $Q_{i,0}$ **OR** $G_i$ **OR** ($W_j$ **AND** TOP[j]) **OR** ($W_i$ **AND** TOP[i])

Where:

- $G_i$ – some Boolean expression, that don't dependent on TOP[i] and TOP[j].

- $W_j$ – some Boolean expression, that don't dependent on TOP[i] and dependent or don't dependent on TOP[j].

- $W_i$ – some Boolean expression, that dependent or don't dependent on TOP[j] and TOP[i] .

For considered combination of the Boolean expressions $Q_{i,j,..}(a[1](t),…, a[n](t))$ the only TOP[i](t) and TOP[j](t) have dual solutions, so all other TOPs have some fixed (single) solutions. If for these solutions and for considered combination of the Boolean expressions $Q_{i,j,..}(a[1](t),…, a[n](t))$ we have, that $G_i(t)$ = **TRUE,** we get that TOP[i](t) = **TRUE** and solution for TOP[i](t) will not be dual. So, $G_i(t)$ = **FALSE.**

Consider some intermediate time moment $t_1$ s.t. $t_0 <= t_1 <= t$, at what some Basic Event has changed his state from **FALSE** to **TRUE** (if all Basic Events did not change its states up time t, TOP[i] also did not change its state, so TOP[i](t) = **FALSE** and we haven't dual solution – see line 1 at the Table 2).

Both all Basic Events and all TOPs could not change its states from **TRUE** to **FALSE,** so boolean expressions $Q_{i,0}$ and $G_i$ also could not change its states from **TRUE** to **FALSE** (because they composed only from operations **AND** and **OR**). So, at the time $t_1$ for $Q_{i,0}$ and $G_i$ there are proved, that $Q_{i,0}(t_1)$ = **FALSE** and $G_i(t_1)$ = **FALSE**, because both $Q_{i,0}(t_0) = Q_{i,0}(t) =$ **FALSE** and $G_i(t_0) = G_i(t) =$ **FALSE**.

At the time $t_1$ the value of the TOP[i] is as early, i.e. TOP[i]($t_1$) = **FALSE**. We don't know values of the $W_i(t_1)$ and $W_j(t_1)$ - they may be both **TRUE** and **FALSE.** But in any case at the time $t_2 = t_1 + \Delta t$ (after some of the Basic Event has changed its state from **FALSE** to **TRUE**) the value of the TOP[i] at this time will be : TOP[i]($t_2$) = $Q_{i,0}(t_1)$ **OR** $G_i(t_1)$ **OR**
{$W_i(t_1)$ **AND** TOP[i]($t_1$) } **OR** {$W_j(t_1)$ **AND** TOP[j]($t_1$) } = **FALSE OR FALSE OR** {$W_i(t_1)$ **AND  FALSE**)} **OR** {$W_j(t_1)$ **AND  FALSE**) } = **FALSE**.

So, after each transformation of the state of any of Basic Events, if before this time moment the states of the TOP[i] and TOP[j] with possible dual solution were **FALSE,** the values of the TOP[i] and TOP[j] are not changed. So TOP[i](t) = **FALSE** and TOP[j](t) = **FALSE.**

Same consideration for situations, when 3 or more TOPs simultaneously have possible dual solutions for some combination of the Boolean expressions

$Q_{i,j,..}(a[1](t),…, a[n](t))$.

## 5. CONCLUSIONS

Early proposed exact analytic methods for calculation of the Arbitrary Fault Tree with Loops are analysed. It is shown, that they don't applicable for Fault Trees with repairable Basic Events, because such Fault Trees can have dual solutions, dependent on pre-history. Otherwise, it is proved, that for Fault Tree with non-repairable Basic Events, which include only gates AND, OR and based of them composed gates (as "K out of M"), the solution may be only uniqueness and so early proposed methods are correct.